\documentclass[pra,aps,showpacs]{revtex4}
\usepackage{amsmath}
\usepackage{epsfig}
\usepackage{array}
\usepackage{natbib}

\begin{document}

\title{\bf An universal algorithm of calculating terms of atomic many-body 
perturbation theory}
\author{V. A. Dzuba}
\affiliation{School of Physics, University of New South Wales, 
Sydney 2052, Australia}
\date{\today}

\begin{abstract}
An algorithm, based on numerical description of the terms of many-body
perturbation theory (Goldstone diagrams), is presented. The algorithm
allows the use of the same piece of computer code to evaluate any particular
diagram in any specific order of the perturbation theory or to calculate
similar terms in other areas of the many-body theory, like e.g. terms in the
coupled-cluster equations.
The use of the algorithm is illustrated by calculating the second and
third order correlation corrections to the removal energies of electrons from
the ground state of sodium, copper and gallium and by calculating the
hyperfine structure constants of sodium in the linearized single-double
coupled cluster approximation.
\end{abstract} 
\pacs{PACS: 31.15.Md,31.25.-v,31.25.Eb}
\maketitle

\section{introduction}

This paper discusses calculations of terms of the many-body theory which arise
when a system of interacting particles is described with the use of a set of
single-particle basis functions. We focus the discussion on the atomic
many-body perturbation theory (MBPT). However, similar approach can be used in
molecular, solid state, nuclear physics, etc. and is not limited to the
perturbation theory. The approach relies of the numerical description of the structure
of the terms (diagrams) and on an universal algorithm which calculates the
terms according to their structure.

Many body perturbation theory (MBPT) is one of the most commonly 
used tools of performing calculations for many-electron systems. 
There is enormous number of examples of its successful application 
to atoms, molecules, etc. (see, e.g.\cite{Lindgren,Wilson}).
However, the use of MBPT has many technical difficulties. 
It is often insufficient to include only lowest-order terms
to obtain desirable accuracy. But inclusion of higher orders leads
to huge increase of the number of terms and, what is even
more important, of the the number of different types of terms, 
or terms which have different structure. Following Goldstone
diagram technique we will use graphic diagrams to show the
structure of the terms. Common programming practice is to write
a separate piece of computer code for each diagram. However,
in higher orders the number of diagrams increases so dramatically
that writing corresponding computer codes becomes a very difficult
task.

There is certainly a need to automatize the use of the
MBPT. This must include at least two stages:
(1) generation of the diagrams, and (2) evaluation of the diagrams.
These two tasks are closely related since diagrams are to be generated in
a form which is then used for their calculation. Generation of diagrams
was discussed in Ref.~\cite{Paldus73,Csepes88,Lyons,Derevianko02,Cannon04,Mathar}. 
We will not discuss it here but rather focus on the diagram evaluation
assuming that the diagrams have unambiguous numerical description.

In the approach used by Derevianko {\em et al}~\cite{Derevianko02,Cannon04}
diagrams are generated together with computer codes for their
calculation. In present paper we present a different approach
which is based on an universal algorithm in which diagram
structure is presented by a set of parameters which are passed to a Fortran
subroutine. This allows the use of a single
piece of computer code to calculate any specific diagram.

We also don't discuss in this paper the task of calculation of the basis of
single-electron states. The algorithm is independent of the basis apart from
the fact that its current implementation assumes the use of basis functions
which correspond to the spherically-symmetric case, i.e their dependence on
angular coordinates is determined by spherical harmonics. The independence on
the basis gives the users flexibility to use a basis which best fits their
needs. We use the B-spline basis~\cite{b-spline}, an alternative choice is
presented in Ref.~\cite{D-basis}. In fact, any functions which constitute a
complete set of states in a spherically-symmetric case can be used.

In this paper we first describe the idea behind the algorithm, the features
of its current implementation, its efficiency and intended use. Then we
describe the algorithm itself. Finally, we present two examples of its use.
First is the many-body perturbation theory calculations of the second and 
third order correlation corrections to the removal energies for an external 
electron in the ground state of sodium, copper and gallium. Second is the
calculation of the hyperfine structure of sodium in the linearized
single-double coupled cluster approximation.

The technique developed in this work was recently used for
the calculation of the relativistic energy shifts of 
frequencies of atomic transitions used in the search for
variation of the fine structure constant in quasar absorption
spectra~\cite{alpha-sd}. The calculations were performed with
the linearized couple-cluster single-double method complemented
by the third-order MBPT. All third-order calculations were performed
using the algorithm presented in this work.

\section{General idea and its implementation}

The algorithm does not represent an independent method of many-body
calculations. It only represents a way of calculating terms (diagrams) in any
many-body method such as MBPT, CC, etc.
Current implementation of the algorithm has certain limitations which will be
discussed below. However, the idea behind it is very general. It can be
formulated in two statements:
\begin{enumerate}
\item The structure of a diagram can be described numerically.
\item Once an unambiguous numerical description of a diagram is given, a
procedure can be build to evaluate the diagram according to its structure.
This procedure can be a very general one, working for any diagram.
\end{enumerate}
This approach can be used in any many-body theory which deals with a system
of interacting particles and uses a set of single-particle basis functions. 
Current implementation has been developed for atomic calculations. It has
features which are natural for atomic calculations and which impose some
limitations on the use of the algorithm. In principle, most of these
limitations can be eased or avoided with appropriate amendments to the
algorithm. These features are:
\begin{itemize}
\item Spherical coordinates are used.
It is assumed that there is a set of 
orthogonal and normalized single-electron basis states which are 
the eigenstates of the Dirac operator in the spherically-symmetric 
case. 
The basis orbitals have standard set of quantum numbers: 
$|i\rangle \equiv |njlm\rangle$, where $n$ is principal
quantum number, $j$ is total momentum, $l$ is angular momentum,
$m$ is projection of $j$. They may come from calculations in
some spherically-symmetric potential, like Hartree-Fock
potential of closed atomic shells, Coulomb nuclear potential,
some parametric potential, etc. They can also be some artificial
functions not directly relevant to any potential. 

The angular dependence of the basis functions is determined by spherical
spinors. This allows the use of standard angular momentum algebra to reduce
integrals over angular variables to the $3j$-symbols and to perform summation
over projections $m$. This procedure, known as {\em angular reduction}, is done
in the algorithm partly analytically and partly numerically. Integrals over
angular variables are still expressed via the $3j$-symbols analytically while
summation over projections $m$ is done numerically. Numerical approach to the
summation over projections allows to perform it in a universal way, so that the
procedure works for any diagram.

In principle, appropriate ammendments to the numerical angular reduction part
of the algorithm could extend it to a non-spherically-symmetric case.


\item In present work we give detailed consideration only to diagrams with two
free ends. These two free ends represent states of a valence electron and
corresponding diagrams represent a correction to the removal energy
of this electron due to correlations with the electrons in atomic core. 
With some minor modification the algorithm can also be used for diagrams with
no free ends (e.g. correlation correction to the energy of atomic core), four
ends (e.g. correlation correction to the Coulomb interaction between two
valence atomic electrons), etc. In fact, we use the later version of the
algorithm for solving of the single-double equations as is discussed in section
\ref{s:SDhfs}. 

\item Present algorithm deals with radial integrals involving two or four
basis functions. The former may represent e.g. matrix elements of an external field
acting on an electron, while the later may represent matrix elements of the
interaction between two electrons. This would cover most of the
applications. However, there are cases when it is not enough. For example,
triple excitation coefficients in the coupled-cluster method have properties
of radial integrals involving six basis functions. Certain modification of the
algorithm may include this case too.

\item The task of calculating single and
double-electron matrix elements is left to the user while
only the rank of the operators is passed to the algorithm
for numerical calculation of angular coefficients.
This gives a lot of flexibility for the algorithm use.
Not only the basis is chosen by the user but also interpretation
of the single and double electron operators is left to the user.
This allows to consider a wide range of problems without making
any adjustments in the algorithm. For example, in most of cases 
the two-electron operator is likely to be an operator of
the non-relativistic Coulomb interaction between electrons.
But, depending on the problem, it can also be Breit relativistic
correction to the inter-electron interaction, specific mass shift
operator, double excitation coefficient in the coupled-cluster calculations,
etc. There is even wider choice for a single-electron
operator. This can be an interaction with external laser field, 
interaction of atomic electrons with electric and magnetic moments
of the nucleus, etc. As a result, the same algorithm can be used
for the MBPT calculations of atomic energy levels, fine and hyperfine 
structure, transition amplitudes, effects of parity non-conservation, 
single excitation coefficient, etc.
\end{itemize}

\section{Efficiency}

The purpose of the algorithm discussed in this paper is to free researcher
time from writing and testing computer codes whenever terms of a many-body
theory need to be calculated. It provides a ready piece of code which can be
used for any terms of any many-body method. In terms of computer resources
needed in most of cases the use of the algorithm would be as efficient as
writing new code. In some special cases when efficiency is important writing
new code might still be a desirable alternative. A code written for specific
diagram may take advantage of its structure to perform the calculations in the
most efficient way. In contrast, the algorithm uses an uniform approach to all
diagrams. 

Computer time needed to calculate a diagram by the algorithm presented in
this paper is given by
\begin{equation}
  T = A c^s v^p,
\label{time}
\end{equation}
where $A$ is a factor which depends on a computer and on the implementation
of the algorithm, $c$ is the number of occupied states, $s$ is the number of
summations over occupied states, $v$ is the number of virtual states, and $p$
is the number of summations over virtual states. It is assumed that all
single-electron states with the same quantum numbers $n,l,j$ but different
projection $m$ have the same radial function. Therefore, the factors $c$ and
$v$ are much smaller than the total number of single-electron states. For
example, the total number of electrons in the cesium core is 54, however the
number of core states with different radial functions is 17, i.e. $c$=17 for
Cs. The value of $v$ depends on the basis. In present work we use the B-spline
basis with 40 B-splines in each partial wave up to $l_{max}=5$ which
corresponds to $v=440-17=423$.

Formula (\ref{time}) suggests that the calculations become unwieldy in higher
orders which have terms with large $p$. In our experience it is impractical to go
beyond $p=4$ unless a supercomputer is used. Terms with $p>4$ first appear in
the forth order of the MBPT. This is rather indicates inefficiency of the MBPT
than inefficiency of the algorithm.
When higher-orders need to be considered the best choice is to use some
all-order technique like, e.g. coupled-cluster (CC) technique 
(see, e.g.~\cite{Lindgren,sd,RCC}) or correlation potential
method~\cite{Dzuba89}, etc. The CC technique includes terms with maximum
number of summations over virtual states. Every iteration of the CC equations
corresponds to the next order of the MBPT but takes exactly the same
time. This means that the CC technique is much more efficient in higher orders
than the MBPT. On the other hand the algorithm can be used to calculate terms
in either of this methods, CC or MBPT alike.

A good example when the algorithm can be extremely useful comes from a
comparison of the CC and MBPT techniques. The CC method misses certain
contributions to the many-body wave function usually starting from the third
order of the MBPT. Sometimes it is important to know the value of missed
terms. The MBPT can be used for this purpose. The main obstacle on this way is
a need to write huge amount of computer code to handle large number of
diagrams in higher orders of the MBPT. In contrast, new new programming is
needed when the algorithm is used.

We stress once more that the algorithms does not represent an
independent method of calculations. It represents a way of evaluating terms in
any many-body method. For example, it can be used to calculate terms in the CC
equations or in MBPT, etc. In most cases it is sufficiently efficient, but in
some cases it is not. Therefore,
the general rule is the following. Give special consideration to the most
computer demanding terms. These terms can be easily identified by large number
of summations over virtual states and there are usually very few of them. 
Then use the algorithm for all other terms.

Note in the end that since the calculation of matrix elements is left to the
user, the efficiency of the algorithm is partly in hands of the user as well. This
cannot change formula (\ref{time}) but can reduce the value of $A$ in it.
For example, with the sufficient computer memory, all single and
double-electron integrals can be calculated in advance and stored in the
computer memory. This would speed up the calculations enormously. 

\section{Algorithm}

\subsection{Numerical representation of the diagrams}

\label{diagr-num}

We use two levels of numerical representation of the diagrams.
One is compact and easy to generate. It uniquely identifies a
diagram but gives insufficient description for its actual 
calculation. A more detailed
description must be created before the diagram can be 
calculated. This is also done automatically.

Following standard diagram technique we distinguish
three types of electron states which are presented on a
diagram by three different types on lines:
\begin{itemize}
\item {\em Core (occupied) state}. This is a state in the atomic core. It is
shown by a line which starts and ends on a vertex and goes from 
right to left as shown by an arrow. It assumes summation over
states of atomic core.
\item {\em Excited (virtual) state}. This is a state above atomic core. It is
shown by a line which starts and ends on a vertex and goes from 
left to right as shown by an arrow. It assumes summation over
all states above the core.
\item {\em Valence state}. This is a state above atomic core for
which matrix element represented by the diagram is calculated.
It is shown by a line which starts or ends on a vertex and has
a free end. It goes from left to right as all other states above the core.
It assumes no summation.
\end{itemize}

\begin{figure}

\begin{tabular}{|c r|c r|}
\multicolumn{1}{m{3cm}}{\epsfig{figure=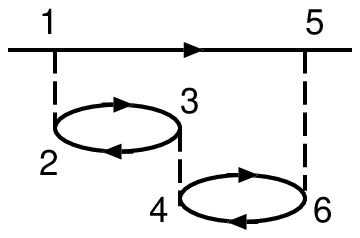,height=2.5cm}}&
\multicolumn{1}{m{4cm}}{\hspace{1cm} 1,5,0,2,3,2,4,6,4} &
\multicolumn{1}{m{3cm}}{\epsfig{figure=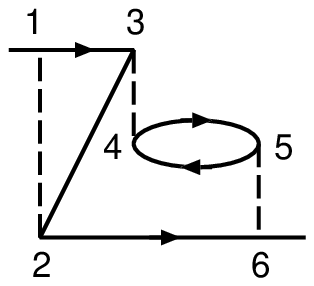,height=2.5cm}}&
\multicolumn{1}{m{4cm}}{\hspace{1cm} 1,3,2,6,0,4,5,4} \\
\multicolumn{1}{m{3cm}}{\epsfig{figure=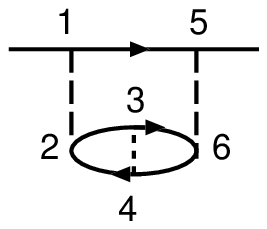,height=2.5cm}}&
\multicolumn{1}{m{4cm}}{\hspace{1cm} 1,5,0,2,3,6,4,2} &
\multicolumn{1}{m{3cm}}{\epsfig{figure=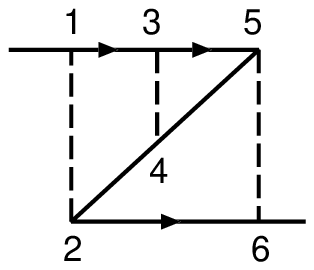,height=2.5cm}}&
\multicolumn{1}{m{4cm}}{\hspace{1cm} 1,3,5,4,2,6} \\
\end{tabular}
\caption{Some third order Goldstone diagrams and their numerical
representation}
\label{f1}
\end{figure}

Compact diagram representation is illustrated on Fig.~\ref{f1}.
It is based on the following rules:
\begin{enumerate}
\item All vertexes are numerated from left to right. Every pair
of vertexes corresponding to a two-electron operator have consequent
numbers (1 and 2, 3 and 4, etc.).
\item The diagram description is created by listing the vertexes in 
order of their appearance while moving along a fermion line in the 
direction shown by an arrow.
\item Fermion line with valence states must be the first in the list.
Zero index is used to separate it from the rest of the diagram.
\item Loops start and end with the same vertex number.
\end{enumerate}

\begin{figure}   
\epsfig{figure=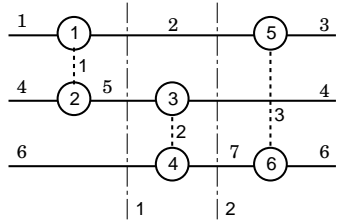,height=3cm}
\caption{Detailed numerical description of the first diagram
from Fig.~\ref{f1}}
\label{f2}
\end{figure}

\begin{figure}   
\epsfig{figure=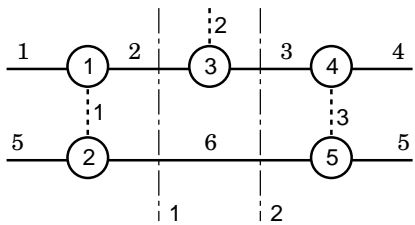,height=3cm}
\caption{Detailed numerical description of a diagram
which contains interaction with external field (singe-electron
operator)}
\label{f3}
\end{figure}

\begin{table}  
\begin{tabular}{|c| c c c|}
\hline
Vertex & Left fermion  & Boson       & Right fermion  \\
number & line number   & line number & line number \\
\hline
\multicolumn{4}{|c|}{Diagram on Fig.~\ref{f2}} \\
\hline
1 & 1 & 1 & 2 \\
2 & 4 & 1 & 5 \\
3 & 5 & 2 & 4 \\
4 & 6 & 2 & 7 \\
5 & 2 & 3 & 3 \\
6 & 7 & 3 & 6 \\
\hline
\multicolumn{4}{|c|}{Diagram on Fig.~\ref{f3}} \\
\hline
 1 & 1 & 1 & 2 \\
 2 & 5 & 1 & 6 \\
 3 & 2 & 2 & 3 \\
 4 & 3 & 3 & 4 \\
 5 & 6 & 3 & 5 \\
\hline
\end{tabular} 
\caption{Vertexes description for diagrams on Fig~\ref{f2} and Fig~\ref{f3}}
\label{tab1}
\end{table}

Fig.~\ref{f2} illustrates detailed numerical description of the 
first digram on Fig.~\ref{f1}. It is generated automatically.
It presents the diagram in a form convenient for the calculations.
Apart from vertexes it also numerate fermion lines (electron states)
and boson lines (interactions). Fermion lines are shown on Fig.~\ref{f2}
as solid horizontal lines while boson lines are dotted vertical lines.
Diagram cross-sections are also shown as dash-dotted lines.
There are closed and open fermion lines. Closed lines start and end 
on vertexes. They assume summation over core or excited states.
Open lines have one free end and correspond to valence states.
They assume no summation. Note that lines 1 and 3 on Fig.~\ref{f2}
correspond to the same valence state. However they are given different
indexes to indicate that there is no summation for these lines. On the 
other hand lines 4 and 6 are closed even though they are shown in two
different pieces each. Table~\ref{tab1} shows that numerical description
of each vertex consists of three whole numbers: left fermion line
number, boson line number and right fermion line number. To complete
the picture the type of every fermion line must be specified: core
state, state above the core or valence state.

Automatic generation of detailed numerical description of a diagram
starting from its compact representation follows the following rules:
\begin{itemize}
\item There is a fermion line between any pair of neighboring vertexes
in the list (see Fig.~\ref{f1}).
\item This line corresponds to a core state if second index is smaller
or to a state above the core otherwise.
\item There is a valence state before the very first index and before
zero index or after the last index if no zero present.
\item Boson lines are always between indexes 1 and 2, 3 and 4, etc.
\item Vertexes, fermion and boson lines are numbered independently.
\end{itemize}

\begin{table}  
\begin{tabular}{ll}
\hline \hline
Step~~ & Result \\
\hline
1 & 1,5,0,2,3,2,4,6,4 \\
2 & 1-f-5-0,\ 2-f-3-f-2,\ 4-f-6-f-4 \\
3 & 1-e-5-0,\ 2-e-3-c-2,\ 4-e-6-c-4 \\
4 & v-1-e-5-v,\ 2-e-3-c-2,\ 4-e-6-c-4 \\
\hline \hline
\end{tabular}
\caption{Generation of the detailed description of the diagram A1.
f is a fermion state (either core or excited), e is an excited state, c
is a core state, v is a valence state.}
\label{tab2}
\end{table}

The process of generating a detailed description of diagram on Fig.\ref{f2}
is illustrated in Table~\ref{tab2}.

Fig.~\ref{f3} shows detailed numerical description of a diagram which
has an interaction with external field (e.g. electric dipole field of
a photon). This is one of the so called {\em structure radiation} diagrams. 
Its numerical description is very similar to those on Fig.~\ref{f2} with
few minor differences: the order of the vertexes is not fixed and
the type of each vertex must be specified (i.e. to which operator
it corresponds). At the moment there is no automatic generation of
this description and it has to be supplied manually.

\subsection{Calculation of the diagrams}

Diagrams differ from each other by number of summations over core and 
excited states, by indexes with enter every single- and double-electron
matrix elements, by energy denominators and by angular coefficients.
An universal algorithm must treat all these differences as parameters
which affect what the program calculates but do not affect its structure.
We will consider them it turn.

\paragraph{\bf Main summation loop.}

The procedure set out in this paragraph is the result of discussions 
with J.S.M. Ginges.

Different diagrams usually have different number of summations over
core and excited states. In a standard approach each summation is
performed in a separate {\em do} loop. Since the number of {\em do}
loops usually cannot be a parameter, here is the need for different 
code for each diagram. In present paper we use a single {\em do}
loop for all summations:
\begin{equation}
  S = \sum_{i=1}^{N_t} s_i.
\label{sum}
\end{equation}
Index of summation $i$ runs over total number
of terms $N_t$ and all other indexes numerating single-electron basis
states are calculated. The total number of terms for any diagram
is given by
\begin{equation}
  N_t = n_{cs}^{n_{cl}} n_{es}^{n_{el}},
\label{nt}
\end{equation}
where $n_{cs}$ is the number of core states, $ n_{cl}$ is the number
of core lines in the diagram, $n_{es}$ is the number of excited
states (states above the core), and $n_{el}$ is the number of lines
in the diagram which represent states above the core. $n_{el}$ does
not include valence lines since they assume no summation.

Calculation of all indexes numerating single-electron states follow
a simple recurrent algorithm. The indexes are ordered and the range
of their change is determined. For every new term in the summation
first index is incremented by one. When it exceeds its maximum value,
it goes back to its minimum value and next index is incremented by one, 
etc.
 
\paragraph{\bf Energy denominators.}

\begin{figure}
\epsfig{figure=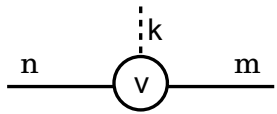,height=1.5cm}
\caption{Vertex corresponding to a single-electron operator.}
\label{me1}
\end{figure}

\begin{figure}
\epsfig{figure=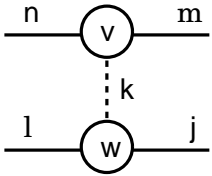,height=2.5cm}
\caption{Vertex corresponding to a double-electron operator.}
\label{me2}
\end{figure}

Total energy denominator of a diagram is calculated as a product of the
energy denominators corresponding to particular diagram cross-sections
\begin{equation}
  \Delta_{total} = \prod_{i=1}^{N_{cs}} \Delta_i.
\label{delta}
\end{equation}
Here $N_{cs}$ is the number of cross-sections for the diagram. 
A cross-section is a vertical line which crosses fermion lines between 
diagram vertexes (see Figs.~\ref{f2} and \ref{f3}).
Recurrent formulas are used to calculate $\Delta_i$.
First
\begin{equation}
  \Delta_0 = 0.
\end{equation}
Then
\begin{equation}
  \Delta_i = \Delta_{i-1} + \epsilon_n - \epsilon_m + \omega,
\label{fme1}
\end{equation}
for a single-electron operator (see Fig.~\ref{me1}) or

\begin{equation}
  \Delta_i = \Delta_{i-1} + \epsilon_n + \epsilon_l - \epsilon_m - \epsilon_j
\label{fme2}
\end{equation}
(see Fig.~\ref{me2}).
Here $\epsilon_i$ is the energy of the
single-electron basis state number $i$, $\omega$ is the frequency
of external monochromatic field ($\omega=0$ for static fields).

\paragraph{\bf Calculation of single- and double-electron matrix elements}

Every term in (\ref{sum}) is a product of single- and double-electron
matrix elements and an angular coefficient. The later will be considered
in next paragraph. As for the product of the single- and double-electron
matrix elements its calculation follows the same pattern as the calculation
of the energy denominators. A transition to every next cross-section of 
a diagram is caused by a singe-electron matrix element or by a 
double-electron matrix element (as on Figs. \ref{me1},\ref{me2}). 
All indexes numerating single-electron
basis states are known on this stage. Therefore, they just passed to
the sections of the program which is assumed to be supplied by the user.

Leaving to the user calculation of the single- and double electron matrix
elements gives a lot of flexibility to the algorithm. 

First, it allows different interpretation of the operators at user 
discretion. For example, in most of cases the double-electron matrix 
element would probably be the matrix element of non-relativistic
Coulomb interaction between electrons. However, it can also be a
matrix element of Breit operator (relativistic correction to the Coulomb
interaction\cite{Breit,BreitDzuba}), or specific mass shift 
operator\cite{sms}, etc.

There is even larger choice for the single-electron operator. This can
be, e.g. electron interaction with nuclear magnetic dipole or nuclear
electric quadrupole moments for calculation of the hyperfine structure
of atomic levels, it can be interaction of atomic electrons with electric
or magnetic laser field for the calculation of transition amplitudes, etc.

The single and double-electron matrix elements can also be interpreted 
as single and double-excitation coefficients in the coupled-cluster 
expansion of the many-electron wave function\cite{sd}. Therefore, the 
algorithm can be used for solving the single-double coupled-cluster
equations.

Second, the user's interpretation of matrix elements allows the use of 
different basis states. For example, B-spline 
basis is used in present work\cite{b-spline}. But it can be a basis of
real Hartree-Fock states, or states calculated in a parametric potential,
Coulomb basis, etc.

Finally, it allows to choose an efficient way of calculating matrix
elements. This is especially important for the double-electron matrix 
elements, since their calculation usually takes most of the computer
time in the MBPT calculations. In the example presented in section
\ref{example} below we calculate Coulomb integrals via Coulomb $Y$-functions 
as it was done in our other works\cite{alpha-sd,core}.
The $Y$-functions are calculated in advance and stored on disk.
The most obvious alternatives to this are to calculate Coulomb
integrals from single-electron wave functions every time they are needed, 
or to calculate them in advance and store on a disk or in computer memory.

\paragraph{\bf Numerical angular reduction.}

Calculation of the angular coefficient consists of two parts. First,
summation over projections and second, calculation of the projection
independent part.

Summation over projections is performed numerically using the same
ideas as for the whole diagram. Summation is done in a single {\em do}
loop which runs over total number of terms which is given by
\begin{equation}
  N_t = \prod_{i=1}^{n_{fl}}(2j_i+1).
\label{nta}
\end{equation}
Here $n_{fl} = n_{cl}+n_{el}$ is the total number of fermion lines 
in the diagram, not counting the valence lines,
$j_i$ is the total momentum of the single-electron state number $i$.
Note that the boson lines don't need to be included into the counting
of the number of terms. This is because corresponding projection is
fixed by the condition $-m_n+q+m_m=0$, where $m_n$ and $m_m$ are
the projections of $j_n$ and $j_m$. Valence lines also don't contribute
to the number of terms (\ref{nta}) because they involve no summation.

Each vertex (see, e.g. Fig.\ref{me1}) is associated with
the $3jm$-symbol
\begin{equation}
(-1)^{j_n-m_n}\left(\begin{array}{ccc} j_n & k & j_m \\
-m_n & q & m_m \end{array} \right).
\label{3jm}
\end{equation}
Here $k$ is the multipolarity of Coulomb interaction or a rank
of single-electron operator.
Every term in the summation over projections is calculated
as a product of $3jm$-symbols corresponding to all vertexes of
the diagram.

Projection-independent part of the angular coefficient corresponding
to Coulomb interaction is calculated as a product of terms
\begin {equation}
(-1)^{j_m+j_j+1}\sqrt{(2j_n+1)(2j_m+1)(2j_l+1)(2j_j+1)}
\left(\begin{array}{ccc} j_n & k & j_m \\
-\frac{1}{2} & 0 & \frac{1}{2} \end{array} \right)
\left(\begin{array}{ccc} j_n & k & j_m \\
-\frac{1}{2} & 0 & \frac{1}{2} \end{array} \right)
\label{Coulomb}
\end{equation}
for every pair of Coulomb vertexes (as on Fig.\ref{me2}).

Calculations of the projection-independent part of a single-electron
operator is left to the user.

\section{Examples}

\subsection{Correlation corrections to the ground-state energies of sodium,
 copper and gallium}

\label{example}
In this section we present the calculation of the energies of the ground
states of sodium, copper and gallium using relativistic Hartree-Fock
method (RHF) and second and third-order of the many-body perturbation 
theory in residual Coulomb interaction. All second and third order 
diagrams are calculated using the technique described above. 

Sodium, copper and gallium represent three different and most common
types of electron structure for atoms with one external electron above
closed shells. MBPT calculations for them is a good illustration on
how the MBPT works for different atomic systems and how it can be
used to significantly improve the accuracy of the calculations.
Calculations for sodium in second and third-order were considered 
before in Ref.\cite{sodium}. This gives us an opportunity to compare 
the results. To the best of our knowledge other two atoms were not
considered before to the third order of the MBPT.

\begin{table}
\begin{tabular}{l l r r r r r}
\hline \hline
Diagram  &  Its numerical & \multicolumn{5}{c}{$\delta E^{(3)}$} \\
as in Ref.~\cite{Blundell90} & representation & Na, 3$s$ & Cu, 4$s$ & 
Ga, 4p$_{1/2}$ & Ga, 4p$_{3/2}$ & Comment \\
\hline
A1 & 1,5,0,2,3,2,4,6,4 &         &       &       &       &             \\
A2 & 1,3,2,6,0,4,5,4   &    180  &  2024 &  1851 &  1815 &    (A1+A2)  \\
A3 & 1,5,0,2,3,6,4,2   &   -352  & -2639 & -3310 & -3260 &             \\
A4 & 1,3,5,4,2,6,0     &     25  &   326 &   692 &   679 &             \\
A5 & 1,5,4,6,0,2,3,2   &    -11  &     8 &  -234 &  -229 &             \\
A6 & 1,3,2,5,4,6,0     &      7  &   101 &   191 &   186 &             \\
A7 & 1,5,3,2,4,6,0     &         &       &       &       &             \\
A8 & 1,3,5,0,2,6,4,2   &   -217  & -1970 & -2369 & -2320 & (A7+A8)   \\

B1 & 5,1,0,2,3,2,4,6,4 &         &       &       &       &           \\
B2 & 5,1,4,2,0,3,6,3   &    -59  &  -196 &  -681 &  -659 & (B1+B2)   \\
B3 & 5,1,0,2,3,6,4,2   &     83  &   290 &  1156 &  1119 &             \\
B4 & 5,1,3,6,4,2,0     &    -32  &  -126 &  -478 &  -469 &             \\
B5 & 5,3,6,2,0,1,4,1   &     25  &    67 &   329 &   320 &             \\
B6 & 5,3,6,1,4,2,0     &     -9  &   -27 &  -127 &  -123 &             \\
B7 & 5,3,1,4,6,2,0     &         &       &       &       &              \\
B8 & 5,3,1,0,2,4,6,2   &     48  &   180 &   580 &   556 &    (B7+B8)  \\

C1 & 3,5,0,1,4,1,2,6,2 &    251  &  2631 &  1523 &  1489 &  2$\times$C1 \\
C2 & 3,5,0,1,4,2,6,1   &    -55  &  -450 &  -489 &  -480 &  2$\times$C2  \\
C3 & 3,1,4,5,0,2,6,2   &         &       &       &       &             \\
C4 & 3,1,5,2,4,6,0     &    -33  &   -50 &  -267 &  -259 &  2$\times$(C3+C4)\\
C5 & 3,5,2,6,0,1,4,1   &    -15  &   -50 &  -219 &  -214 &  2$\times$C5     \\
C6 & 3,5,1,4,2,6,0     &      7  &    48 &    90 &    88 &  2$\times$C6     \\
C7 & 3,1,4,5,2,6,0     &         &       &       &       &             \\
C8 & 3,1,5,0,2,4,6,2   &    -33  &  -133 &  -174 &  -170 &  2$\times$(C7+C8)\\

D1 & 5,3,0,1,4,1,2,6,2 &   -137  &  -560 &  -807 &  -784 &  2$\times$D1     \\
D2 & 5,3,0,1,4,2,6,1   &     34  &   115 &   245 &   238 &  2$\times$D2     \\
D3 & 5,3,1,4,0,2,6,2   &         &       &       &       &             \\
D4 & 5,3,1,6,2,4,0     &     97  &   378 &   569 &   544 &  2$\times$(D3+D4)\\
D5 & 5,1,6,3,0,2,4,2   &     83  &   105 &   296 &   290 &  2$\times$D5     \\
D6 & 5,1,3,2,6,4,0     &    -12  &   -37 &  -111 &  -108 &  2$\times$D6     \\
D7 & 5,1,6,3,2,4,0     &         &       &       &       &             \\
D8 & 5,1,3,0,2,6,4,2   &     65  &   259 &   534 &   510 &  2$\times$(D7+D8) \\

E  & 1,3,5,0,2,4,6,2   &    237  &  1913 &  3098 &  3045 &    (E1+E2) \\
F  & 5,3,1,0,2,6,4,2   &    -54  &  -220 &  -606 &  -582 &    (F1+F2) \\
G  & 5,1,3,0,2,4,6,2   &    -75  &  -349 &  -730 &  -700 &  2$\times$(G1+G2) \\
H  & 3,1,5,0,2,6,4,2   &     26  &    54 &    28 &    27 &  2$\times$(H1+H2) \\
I1 & 3,0,1,5,4,1,2,6,2 &         &       &       &       &             \\
I2 & 3,1,5,4,0,2,6,2   &  -4692  &-22898 &-19732 &-19486 &    (I1+I2)  \\
I3 & 3,0,1,6,2,5,4,1   &         &       &       &       &              \\
I4 & 3,1,5,2,6,4,0     &   1509  &  5361 &  5902 &  5833 &    (I3+I4) \\
J1 & 3,0,1,4,5,1,2,6,2 &   4603  & 21668 & 19018 & 18807 &    (J1+J2) \\
J3 & 3,0,1,4,5,2,6,1   &  -1480  & -5072 & -5644 & -5584 &    (J3+J4) \\
K1 & 5,0,1,6,3,1,2,4,2 &   -290  & -5272 &  -895 &  -862 &  2$\times$(K1+J2) \\
K3 & 5,3,1,6,0,2,4,2   &    -55  & -1666 &  -275 &  -264 &  2$\times$(K3+J4) \\
L1 & 5,0,1,3,6,1,2,4,2 &    267  &  3181 &  1046 &  1009 &  2$\times$(L1+J2) \\
L3 & 5,1,3,6,0,2,4,2   &     28  &   953 &   230 &   230 &  2$\times$(L3+J4) \\
\hline                                                  
\multicolumn{2}{c}{Total} & -82 & -2054 &    826 &   830 &  \\
\hline \hline
\end{tabular}
\caption{Third order contributions to the ground state energies of sodium,
copper and gallium (cm$^{-1}$).}
\label{E3}
\end{table}

We use the $V^{N-1}$ approximation, which is the standard approach
for atoms with one valence electron. Initial Hartree-Fock procedure
is performed for a closed-shell ion, with the valence electron removed.
Then, the states of the valence electron are calculated in the field
of frozen core. We use the B-spline technique\cite{b-spline} to
construct a full set of single-electron orbitals. We use 40 B-splines
of order $k=9$ in each partial wave up to maximum angular momentum
$l_{max}=5$. Single-electron basis functions are calculated as linear
combination of B-splines in a cavity of radius $R_{max} = 40 a_B$
which are eigenstates of the relativistic Hartree-Fock Hamiltonian.

Calculations of the second-order correlation corrections are relatively
simple and were considered many times before 
(see, e.g.\cite{Dzuba83,Johnson87,sodium}).
We will not discuss them here focusing rather on the third-order corrections.
Third-order correlation corrections for the energies of the single-valence
electron atoms were discussed before in Refs.\cite{Blundell88,Blundell90}.
The total number of the third order diagrams is 76. This can be reduced to 52
if symmetry conditions are used. This can be further reduced to 12 
if direct and exchange terms are grouped together. We follow 
Ref.\cite{Blundell90} to name these 12 terms by letters from A to L.
The results of calculations are presented in Table~\ref{E3} separately
for most of the 52 diagrams. 

All diagrams are calculated with the same piece of code using numerical
description of the diagrams as presented in the second column of the table.
Some diagrams are grouped together with their exchange companions.
This is indicated in the {\em comment} column of the table. For example,
diagram A2 is an exchange companion of diagram A1. They were calculated 
together and only the sum is presented in the table. This is indicated
as the A1+A2 comment in the table. Diagrams C,D,G,H,K and L have complex
conjugated companions with exactly the same value. This can be taken
into account by multiplying the diagrams by the factor of 2 as indicated
in the table.

There is strong cancellation between contributions from different diagrams,
the strongest is for diagrams I and J 
(see also Refs.\cite{Blundell88,Blundell90}).
The final answer for the total third-order correction is much smaller
than the largest contributions. However, the final answer is stable
and not very much sensitive to numerical uncertainty. This is because all
diagrams are calculated in a very similar way which leads to cancellation
of numerical error.

\begin{table}
\begin{tabular}{l r r r r}
\hline \hline
      & \multicolumn{1}{c}{Sodium}
      & \multicolumn{1}{c}{Copper}
      & \multicolumn{2}{c}{Gallium} \\
      & \multicolumn{1}{c}{3s}
      & \multicolumn{1}{c}{4s}
      & \multicolumn{1}{c}{4p$_{1/2}$}
      & \multicolumn{1}{c}{4p$_{3/2}$} \\
\hline                                         
RHF   & -39951 & -52302 & -43033 & -42294 \\   
E2    &  -1277 &  -7607 &  -6404 &  -6280 \\   
E3    &    -82 &  -2054 &    826 &    830 \\   
Total & -41310 & -61963 & -48611 & -47744  \\  
\hline
Experiment\cite{Moore}
      & -41450 & -62317 & -48380 & -47554 \\   
\hline \hline
\end{tabular}
\caption{Ground state energies of sodium,
copper and gallium (cm$^{-1}$) in RHF approximation, second-order
correlation corrections(E2) and third-order correlation correction(E3); 
comparison with experiment.}
\label{E2E3}
\end{table}

Table~IV 
summarize the results of the calculations for the
ground-state energies of Na, Cu and Ga in the RHF approximation and 
with inclusion of the second-order (E2) and third-order (E3) 
correlation corrections. For gallium we include both components of the
fine structure doublet. The sum of all contributions (RHF+E2+E3) is 
compared with experiment.

The analysis of the correlations show some differences as well as
similarities between the elements. The total value of the correlation
correction which is the difference between the RHF and experimental 
values (we neglect Breit and QED corrections here) is 3.6\% for Na, 
15\% for Cu and 11\% for Ga. The correlations are strongly dominated
by the second order as it is well known from a number of calculations
(see, e.g.\cite{Dzuba83,Blundell88,Blundell90}). 
However, the third-order correlation correction
is not small and its inclusion leads to further significant improvement
of the results for all three atoms. This is especially prominent for
the case of copper where the third-order correction is only four times
smaller that the second-order correction and constitutes about 3\% of
the experimental value. However, the final result for copper is within 1\%
of the experiment.

Fine structure interval of the $4p$ ground state of gallium is also
significantly improved when second and third-order correlation corrections 
are included (see Table~IV).

The results for sodium presented in this section are in good agreement
with previous calculations\cite{sodium}.

\subsection{Calculation of the hyperfine structure of sodium in the SD
  approximation}

\label{s:SDhfs}
It is known that many properties of light alkaline atoms can be described to
high precision within the linearized single-double coupled cluster
approximation (SD) (see, e.g. Ref.~\cite{Blundell89}). In this approximation
the many-electron wave function of an atom is written as an expansion over 
single and double excitations from the reference Hartree-Fock wave functions.
Expansion coefficients $\rho$ satisfy the set of self-consistent
equations. The SD equations for the excitations from the atomic core have the
form   
\begin{eqnarray}
  &(\epsilon_a - \epsilon_m)\rho_{ma} = \sum_{bn}\tilde g_{mban}\rho_{nb} + 
 \sum_{bnr}g_{mbnr}\tilde\rho_{nrab}- \sum_{bcn}g_{bcan}\tilde\rho_{mnbc},
 \nonumber \\
  &(\epsilon_a+\epsilon_b-\epsilon_m-\epsilon_n)\rho_{mnab} = g_{mnab}+ \sum_{cd}g_{cdab}\rho_{mncd}+
  \sum_{rs}g_{mnrs}\rho_{rsab} \\
& + \left[ \sum_r g_{mnrb}\rho_{ra}-\sum_c g_{cnab}\rho_{mc} 
+\sum_{rc}\tilde g_{cnrb}\tilde \rho_{mrac} \right]+ \left[ \begin{array}{ccc}
 a & \leftrightarrow & b \\
 m & \leftrightarrow & n \end{array} \right]. \nonumber
\label{lcore}
\end{eqnarray}
Here parameters $g$ are Coulomb integrals 
\[ g_{mnab} = \int \int \psi_m^\dagger(r_1) \psi_n^\dagger(r_2)e^2/r_{12}
\psi_a(r_1)\psi_b(r_2)d\mathbf{r}_1d\mathbf{r}_2, \] 
parameters $\epsilon$ are the single-electron Hartree-Fock
energies. Coefficients $\rho$ are to be found by solving the equations
iteratively starting from
\[ \rho_{mnij} = \frac{g_{mnij}}{\epsilon_i+\epsilon_j-\epsilon_m-\epsilon_n}. \]
Indexes $a,b,c$ numerate states in atomic core, indexes
$m,n,r,s$ numerate states above the core, indexes $i,j$ numerate
any states.

The SD equations for a particular valence state $v$ can be obtained from
(\ref{lcore}) by replacing index $a$ by $v$ and replacing $\epsilon_a$ by
$\epsilon_v + \delta\epsilon_v$ where
\[  \delta \epsilon_v = \sum_{mab}g_{abvm}\tilde\rho_{mvab}+
  \sum_{mnb}g_{vbmn}\tilde\rho_{mnvb} \]
is a correction to the energy of the valence electron.
The SD equations are solved iteratively first for the core and than for as many  
valence states $v$ as needed. The hyperfine structure (hfs) constants are found as
an expectation values of the hfs operator over the SD wave functions.

Here we would like to demonstrate that the terms in the SD equations
(\ref{lcore}) as well as the expressions for the matrix elements (see below)
can have numerical representation which opens a way for their evaluation using
the algorithm described above. Before doing so let's note that computation
time for solving the SD equations is strongly dominated by singe term
containing double summation over virtual states 
\begin{equation}
 \sum_{rs}g_{mnrs}\rho_{rsab} . 
\label{monster}
\end{equation}
Therefore, it can be coded separately, using the most efficient way of its
evaluation. All other terms can be treated by the universal algorithm. This is
very general approach to the most efficient use of the algorithm: give special
attention to the most computationally demanding terms (there are usually very
few of them), use the algorithm for everything else.

Note also, that in contrast to the standard MBPT expressions the terms in the
SD equations (\ref{lcore}) have no energy denominators. Therefore, the
algorithm should be modified to remove calculations of the energy
denominators. This makes it even simpler.

We use whole numbers as specified in Table~\ref{mn12} for the numerical
description of the terms in (\ref{lcore}). Each of the values ($\rho$ or $g$)
are described as a sequence of three (for $\rho_{ij}$) or five (for
$\rho_{ijkl}$ or $g_{ijkl}$) whole numbers. First of these numbers indicate the
type of the value ($\rho$, $\tilde\rho$, $g$ or $\tilde g$), other numbers
are the indexes. Negative numbers are used for occupied states and positive
numbers are used for virtual or valence states. Summation is assumed over
repeated indexes. Zero is used as a delimiter. Numerical representation of the
SD terms (apart from the (\ref{monster}) is given in Table~\ref{tb:sd}. 
This representation can be easily translated into a detailed diagram
description discussed in section~\ref{diagr-num}.

Once the SD equations are solved for the core and for the valence states of
interest, the hfs constants for these valence states can be calculated as
expectation values of the hfs operator over the SD vale
function. Corresponding expressions can be found in Ref.~\cite{Blundell89} and
Table~\ref{sd-hfs}. We also present in the Table the numerical representation
of these terms. It is very similar to the case of the SD terms with one small
difference. Now we have to distinguish between single excitation coefficients
$\rho{ij}$ and single-electron matrix elements of the hfs operator. We use
indexes 1 and 2 for this purpose.

The results presented in the Table are in good agreement with the calculations
of Ref.~\cite{Safronova99} and with experiment.

\begin{table}
\begin{tabular}{c c l}
\hline \hline
Notation  &  Its numerical  & Comment \\
          & representation  &  \\
\hline
$\rho$       & 1 & $\rho_{ij}$ or $\rho_{ijkl}$ \\
$\tilde\rho$ & 2 & $\tilde\rho_{ijkl}=\rho_{ijkl}-\rho_{ijlk}$ \\
$g$          & 3 & Coulomb integral $g_{ijkl}$ \\
$\tilde g$   & 4 & $\tilde g_{ijkl}=g_{ijkl}-g_{ijlk}$ \\
$a,b, \dots$ & -1,-2, $\dots$ & Occupied (core) states \\
$m,n, \dots$ & 1,2, $\dots$ & Virtual or valence states \\
\hline
\end{tabular}
\caption{Numerical representation of the parameters and variables of the SD equations.}
\label{mn12}
\end{table}

\begin{table}
\begin{tabular}{ll}
\hline \hline
Term      &  Its numerical representation \\
\hline
$ g_{cdab}\rho_{mncd}$ & 3,-3,-4,-1,-2,0,1,1,2,-3,-4 \\
$ g_{mnrb}\rho_{ra}$  & 3,1,2,3,-2,0,1,3,-1 \\
$ g_{cnab}\rho_{mc}$  & 3,-3,2,-1,-2,0,1,1,-3 \\
$ \tilde g_{cnrb}\tilde\rho_{mrac}$ & 4,-3,2,3,-2,0,2,1,3,-1,-3 \\
\hline
\end{tabular}
\caption{Numerical representation of the terms of the SD equations}
\label{tb:sd}
\end{table}

\begin{table}
\begin{tabular}{l l l r r}
\hline \hline
\multicolumn{2}{c}{SD terms}  &  Their numerical  &
\multicolumn{1}{c}{$3s_{1/2}$} & \multicolumn{1}{c}{$3p_{1/2}$} \\
\multicolumn{2}{c}{as in Ref.~\cite{Blundell89}} & representation & & \\
\hline
    & $ z_{vw}$ &  & 623.75 & 63.43 \\   
$a$ & $ z_{am}\tilde\rho_{wmva}+c.c.$                  & 2,-1,1,0,2,2,1,3,-1                 & 122.43 &  15.34 \\
$b$ & $-z_{av}\rho_{wa}+c.c.$                          & 2,-1,1,0,1,2,-1                     &  16.48 &   2.45 \\
$c$ & $ z_{wm}\rho_{mv}+c.c.$                          & 2,2,1,0,1,1,3                       & 104.94 &  10.17 \\
$d$ & $ \rho_{mv}z_{mn}\rho_{nv}$                      & 1,3,1,0,2,1,2,0,1,2,4               &   4.41 &   0.43 \\
$e$ & $ \rho_{vb}z_{ab}\rho_{wa}$                      & 1,1,-1,0,2,-1,-2,0,1,-2,2           &   0.11 &   0.03 \\
$f$ & $-z_{av}\rho_{mw}\rho_{ma}+c.c.$                 & 2,2,-1,0,1,-1,1,0,1,1,3             &   0.78 &   0.10 \\
$g$ & $-z_{mv}\rho_{ma}\rho_{wa}+c.c.$                 & 2,2,1,0,1,1,-1,0,1,-1,3             &   0.01 &   0.01 \\
$h$ & $ \rho_{nw}\tilde\rho_{nmva}z_{am}+c.c.$         & 2,3,-1,1,2,0,1,1,4,0,2,2,-1         &   8.07 &   0.86 \\
$i$ & $ \rho_{ma}z_{mn}\tilde\rho_{wnva}+c.c.$         & 2,3,-1,4,1,0,2,1,2,0,1,2,-1         &   0.06 &   0.05 \\
$j$ & $-\rho_{mb}\tilde\rho_{wmva}z_{ab}+c.c.$         & 2,2,-1,3,1,0,1,1,-2,0,2,-2,-1       &   0.27 &   0.09 \\
$k$ & $ \rho_{vb}\tilde\rho_{wmab}z_{am}+c.c.$         & 1,2,-2,0,2,-1,-2,3,1,0,2,1,-1       &   0.94 &   0.11 \\
$l$ & $-z_{av}\rho_{mb}\tilde\rho_{wmab}+c.c.$         & 2,2,-1,0,1,1,-2,0,2,-1,-2,3,1       &  -0.41 &  -0.07 \\
$m$ & $-z_{mv}\rho_{na}\tilde\rho_{nmwa}+c.c.$         & 2,3,1,0,2,1,2,-1,4,0,1,-1,2         &  -1.35 &  -0.10 \\
$n$ & $ z_{ab}\rho_{mnwb}\tilde\rho_{nmva}$            & 2,-1,-2,0,2,3,-2,1,2,0,1,1,2,4,-1   &   0.96 &   0.30 \\
$o$ & $ \tilde\rho_{vmbc}z_{ab}\tilde\rho_{wmac}$      & 2,2,1,-1,-3,0,2,-1,-2,0,2,-2,-3,3,1 &   4.12 &   0.65 \\
$p$ & $ \tilde\rho_{rmwa}z_{mn}\tilde\rho_{rnva}$      & 2,4,-1,3,1,0,2,1,2,0,2,3,2,5,-1     &  10.68 &   0.81 \\
$q$ & $ z_{mn}\rho_{vmab}\tilde\rho_{wnba}$            & 2,1,2,0,1,3,2,-1,-2,0,2,-1,-2,4,1   &   1.61 &   0.08 \\
$r$ & $-\tilde\rho_{vnwb}z_{am}\tilde\rho_{nmab}+c.c.$ & 2,3,1,4,-2,0,2,-2,-1,2,1,0,2,2,-1   &   3.47 &   0.44 \\
$s$ & $ z_{av}\rho_{mnwb}\tilde\rho_{nmab}+c.c.$       & 2,3,-1,0,2,-1,-2,1,2,0,1,1,2,4,-2   &  -7.19 &  -0.75 \\
$t$ & $-z_{mv}\rho_{nmab}\tilde\rho_{wnba}+c.c.$       & 2,3,1,0,1,1,2,-1,-2,0,2,-1,-2,4,2   &  -7.88 &  -1.02 \\
\hline
\multicolumn{2}{c}{Sum $a$ to $t$}                    &
& 262.51 & 29.97 \\

\multicolumn{2}{c}{Normalization}                     & &       1.004 & 1.001 \\
\multicolumn{2}{c}{Final        }                     & &  882.85 & 93.26 \\
\multicolumn{2}{c}{Experiment\cite{Happer,Wij}}       & &  885.8~ & 94.44(13) \\
\hline \hline
\end{tabular}
\caption{Hyperfine structure constants for $^{23}$Na (MHz) in the SD approximation.}
\label{sd-hfs}
\end{table}

\section{Conclusions}

We present an universal algorithm of calculating terms of the many-body
perturbation theory. The algorithm assumes the basis of spherically 
symmetric relativistic single-electron states with the $n,j,l,m$
quantum numbers. Apart from this and available computer power the 
algorithm is practically free from any limitations. It can be used 
to calculate any diagram of any order of MBPT with any set of
single and double electron operators.

The algorithm presents a way to go to higher-orders of MBPT when
sufficient computer power is available. It can also be used for 
solving single-double coupled-cluster equations or for any other
calculations involving terms similar to those of the MBPT.

Calculations of the second and third order correlation corrections
to the removal energies of electrons from the ground-states of sodium, 
copper and gallium as well as the hyperfine stricture constants of sodium in
the SD approximation demonstrated the use of the technique. The total number
of terms handled by the algorithm in these calculations is about one hundred  
and the calculations would be very difficult without it.

\begin{acknowledgments}

The author is grateful to J.S.M. Ginges, S. Blundell, and W. R. Johnson 
for useful discussions.
The work is partly supported by the Australian Research Council.

\end{acknowledgments}

\end{document}